\documentclass[a4paper,11pt]{article}
\usepackage{pos}
\usepackage{wrapfig}

\title{Performance study update of observations in divergent mode for the Cherenkov Telescope Array}
 \ShortTitle{Divergent Pointing Mode for CTAO}

\author*[a]{A. Donini}
\author[b,c]{I. Burelli}
\author[d]{O. Gueta}
\author[e,c]{F.Longo}
\author[d]{E.Pueschel}
\author[d]{D. Tak}
\author[b,c]{A. Vigliano}
\author[g]{T. Vuillamme}
\author[f]{O. Sergijenko}
\author[h]{A. Sarkar}

\affiliation[a]{INAF - Osservatorio Astronomico di Roma, Italy}
\affiliation[b]{University of Udine, Italy}
\affiliation[c]{Istituto Nazionale di Fisica Nucleare (INFN) Trieste, Italy}
\affiliation[d]{Deutsches Elektronen-Synchrotron, Zeuthen, Germany}
\affiliation[e]{University of Trieste, Italy}
\affiliation[g]{Univ. Savoie Mont Blanc, CNRS, Laboratoire d'Annecy de Physique des Particules - IN2P3, France}
\affiliation[f]{Astronomical Observatory of Taras Shevchenko National University of Kyiv}
\affiliation[h]{University of Oxford, United Kingdom}

\onbehalf{for the CTA Consortium and the CTA Observatory} 

\abstract{Due to the limited field of view (FoV) of Cherenkov telescopes, the time needed to achieve target sensitivity for surveys of the extragalactic and Galactic sky is large. To optimize the time spent to perform such surveys, a so-called “divergent mode” of the Cherenkov Telescope Array Observatory (CTAO) was proposed as an alternative observation strategy to the traditional parallel pointing.
In the divergent mode, each telescope points to a position in the sky that is slightly offset, in the outward direction, from the original center of the field of view. This bring the advantage of increasing the total instantaneous arrays' FoV. From an enlarged field of view also benefits the search for very-high-energy transient sources, making it possible to cover large sky regions in follow-up observations, or to quickly cover the probability sky map in case of Gamma Ray Bursts (GRB), Gravitational Waves (GW), and other transient events.
In this contribution, we present the proposed implementation of the divergent pointing mode and its first preliminary performance estimation for the southern CTAO array.}

\ConferenceLogo{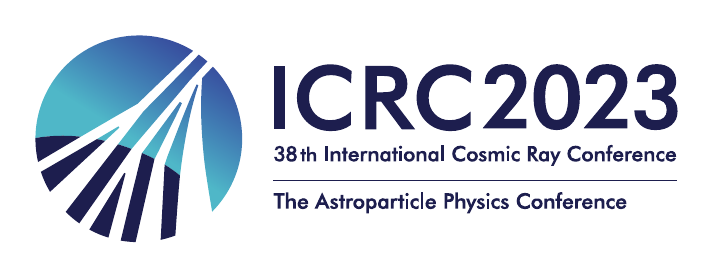}

\FullConference{%
38th International Cosmic Ray Conference (ICRC2023)\\
  26 July - 3 August, 2023\\
  Nagoya, Japan}

\begin{document}
\maketitle

\section{Introduction}
The Cherenkov Telescope Array Observatory (CTAO) is going to be the major next-generation observatory for ground-based very-high-energy gamma-ray astronomy \cite{ref:cta}. A significant improvement in angular resolution, energy resolution and sensitivity with respect to existing IACT experiments will be achieved by building a large number of telescopes with three different sizes across two sites, one in the Northern Hemisphere, at La Palma in the Canary Islands, and one in the Southern Hemisphere, at Cerro Paranal in Chile. The Small-Sized Telescopes (SSTs) will have a primary mirror of about 4 meters in diameter, the Medium-Sized Telescope (MSTs) with a primary mirror of 12 meters and the Large-Sized Telescopes (LSTs) that will be the largest one, with a 23 meters primary mirror.
The work, here presented, shows for the first time the performances of different divergent pointing configurations for the Southern CTA site. The dataset used is based on Monte Carlo (MC) simulations tailored to each divergent configuration.

\section{Divergent pointing}

    Divergent pointing was first introduced as a possible pointing strategy for CTAO to optimize the extragalactic survey task by Dubus et al. in 2013 \cite{ref:dubus2013}. This study analysed the behavior of an array of H.E.S.S.-like MSTs arranged to cover a region of the sky of $20^\circ\times20^\circ$.
    \begin{wrapfigure}{r}{5cm}
        \centering
        \includegraphics[width=4.7cm]{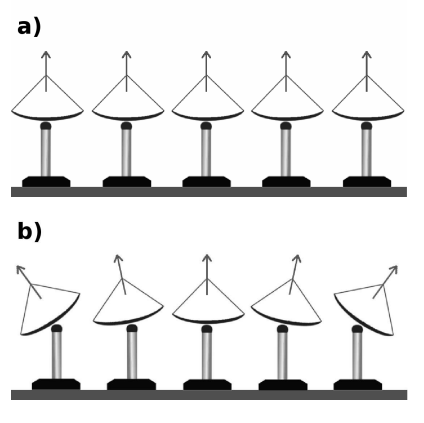}
        \caption{Two modes of configuration of the telescope system: a) normal (parallel) mode; b) divergent mode. From \cite{ref:Szanecki2015}}
        \label{fig:poinitngs}
    \end{wrapfigure}
    The idea underlying the divergent mode is to slightly incline each telescope into the outward direction by an angle increasing with the telescope distance from the center of the array as shown in \figurename{ \ref{fig:poinitngs}}. \\
    The advantage of this configuration is an increased FoV, which reduces the time needed to cover large areas of the sky and also increases the probability to observe a transient source inside this enlarged FoV. The drawback is that reduced energy and angular resolution are expected.
    The goal should be to maximize the size of the field of view while maintaining a good performance. However, different divergent pointing configurations are possible, and the most suitable for the science goals one intends to achieve should be chosen.
    Following the work presented by Dubus et al., Szanecki et al.\cite{ref:Szanecki2015} analysed the performance of an array of MSTs showing that, for the configuration considered, divergent pointing could lead to a gain of a factor 2.3 in time for the extragalactic survey for a defined sensitivity level. A more recent study  \cite{ref:PhD_th_Donini}, analysed the performance of the CTAO Northern array in the so called Omega or Baseline configuration (4 LSTs + 15 MSTs) using the simulations from the third massive MC production (\texttt{prod3}\cite{ref:prod3}). The configurations studied were optimized for the northern site, increasing the array FoV of a variable factor going for 1.5 to 5. \\
    The need to cover a large area of the sky is a requirement not only for the extragalactic survey. Since Gravitational Waves are poorly localized, the search for their electromagnetic counterparts is performed inside a large uncertainty region ($\sim$ 100-1000 deg$^2$), making it a topic that can benefit from the divergent  pointing strategy. 
    The same can be applied to GRBs which are poorly localized. 
    As mentioned before, an enlarged FoV means also an enhanced probability for a transient source to fall inside the region of the sky observed by the instrument. This could hopefully lead to the observation of the onset phase of Gamma-Ray Bursts, also known as prompt emission, which has never been observed with IACTs.\\

\subsection{The \texttt{divtel} library}
\begin{figure}[h!]
    \centering
        {\includegraphics[width=6cm]{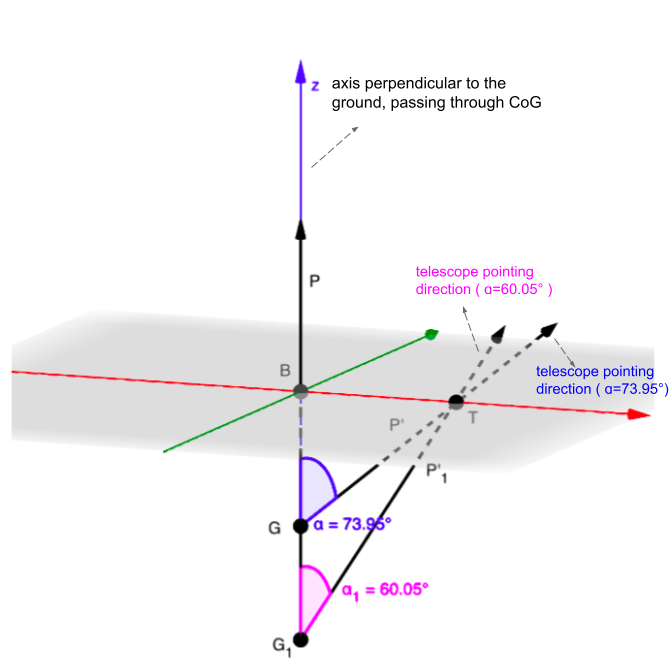}}
         \hspace{5mm}
        {\includegraphics[width=6cm]{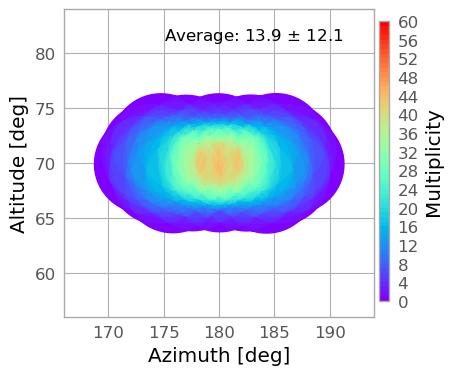}}
        \caption{\textbf{Left:} Concept used to define the divergent pointing from a parallel one for a telescope. \textbf{B} represents the ground position of the array CoG, \textbf{T} is the telescope ground position. \textbf{G} and \textbf{G'} represent the ground point for different divergent configuration. This point is defined in order to control the single telescope pointing direction defined by the lines \textbf{P} and \textbf{P'}. 
        \textbf{Right:} geometric FoV of divergent pointing cfg2. Color codes the telescope multiplicity across the FoV whose average value is printed in the upper part. }
        \label{fig:pointing_tool}
    \end{figure}

    The main code that calculates the optimal pointing direction for each telescope is hosted under the official CTAO GitHub page \cite{ref:divtel} and should be used by the Array Control And Data Acquisition system (ACADA) group, which takes care of the implementation of the divergent pointing strategy in the control system of CTAO.
    The main idea of the algorithm is to have a single parameter, called \texttt{div}, through which the divergent configuration of the array can be modified.
    However, as for now, this implementation does not allow to control the radial symmetry of the FoV and should be soon updated to have better control of the final geometry of the chosen configuration.
    The current version of the code draws an imaginary line aligned with the telescope pointing direction and connecting each telescope ground position, represented by \textbf{T} in \figurename{ \ref{fig:pointing_tool}} (left), to an axis perpendicular to the ground and passing through the Center of Gravity (CoG) of the array, labeled with z in the same fig.
    These lines are defined so as to meet all in the same point, called ground point (\textbf{G}), whose position along the z axis determines the pointing direction of single telescopes.
    In fact the parameter \texttt{div} is directly related to the angle between the axis of the CoG and the telescope pointing direction by the equation 
    
    \begin{equation}
    \begin{split}
        div&=\sin(\alpha)   \\
        |\overrightarrow{GB}|&=\frac{f}{\tan({\arcsin({div}))}}
    \end{split}
    \end{equation}
    where f is a normalization factor taking into account the real distance of the telescope from \textbf{B}. 
    Notice that our goal is to define $|\overrightarrow{GB}|$, which can be defined from the div (or $\alpha$) value of a single telescope. Once the position of the ground point is defined the array pointing directions can be computed. A useful analogy already introduced in previous works is the one of an umbrella where moving the runner up and down the inclination of the stretchers is modified accordingly.
    
\section{Simulations and data analysis}
    The response of the telescope array to Extensive Air Showers (EAS) induced by gamma-rays and proton background is simulated thanks to two packages: \texttt{CORSIKA} and sim\_telarray\cite{ref:corsika}. Both packages are the standard for CTAO simulations. The former takes care of simulating the development of the shower in the atmosphere, the latter simulates the array response to the shower.
    For this contribution gamma-rays have been simulated both for a point-like source for an incoming direction of Zd=20$^\circ$ and Az=0$^\circ$ and for a diffuse component with a incoming directions isotropically distributed inside a cone of aperture 20$^\circ$ centered around the same direction of the point-like gammas. This component is used both to train the random forests and to compute the array response to a diffuse source. The parameters related to the array, such as telescope ground positions and camera and telescope types are taken from an updated version of the \texttt{Prod5} model \cite{ref:prod5}. The simulated array configuration contains globally 87 telescopes, of which only 60 are used - 4 LSTs, 14 MSTs and 42 SSTs - in order to have the analysis of a subarray more similar to what is called Alpha configuration. The main difference is the addition of the LSTs to the configuration. This class of telescopes is not included in the Alpha configuration but since funding for at least two of them have been allocated we added them to our simulations. 
    
    \begin{table}[h!]
        \centering
        \begin{tabular}{|c c c c c|} 
             \hline
             Cfg name & div  & FoV (deg$^2$)  &  FoV$_{eff}$ (deg$^2$) & m$_{ave}$\\  
             \hline
             parallel & 0.0 & 62 & 62  & 53.4\\
             cfg1.5 & 0.0022 & 142 & 127 & 23.4 \\ 
             cfg2 & 0.0043 & 238 & 198 & 13.9\\
            cfg3 & 0.008 & 452 & 334& 7.3 \\
             cfg4 & 0.01135 & 688  & 446 & 4.8 \\
             cfg5 & 0.01453 & 942 & 531 & 3.5 \\ [1ex] 
             \hline
        \end{tabular}
        \caption{Hyper FoV and average multiplicity of the divergent configurations simulated for site south. The first line reports the corresponding values for parallel pointing, as a reference. FoV$_{eff}$ is the area of the FoV simultaneously observed by at least three telescopes. $m_{ave}$ is the average telescope multiplicity, defined as the average number of telescopes covering the FoV.}
        \label{table:divergent_details}
    \end{table}
   
    Divergent pointing simulations require that only the used telescopes have to be simulated. The reason is that the pointing direction of single telescopes is defined starting form the array Center of Gravity (CoG), which changes with the configuration. Nonetheless, the Southern site has a quite symmetric distribution of the telescopes and the difference between the simulated array CoG and the analysed one is negligible (6m).
    The energy range and number of particles that were simulated are consistent with the parallel production \cite{ref:sim_details}. Both point-like and diffuse gamma-rays have been simulated. The latter are used to train the energy regression model and the particle classification one, while point-like gammas are used to produce IRFs for point-like sources.\\
    The \texttt{div} values considered in this study are the same defined in \cite{ref:PhD_th_Donini} and they are reported in table \ref{table:divergent_details} together with other relevant values of the simulated configurations. FoV represents the global, geometric Field of View of the configuration while FoV$_{eff}$ represents the area geometrically covered by at least three telescopes. At these stage no trigger information are available so the information on FoV$_{eff}$ is only geometrical. \figurename{ \ref{fig:pointing_tool}} (right) shows as example the geometric FoV and multiplicity for cfg2. The choice to have at least three telescopes is arbitrary but we think it might be a good starting point in order to guarantee a proper shower reconstruction. The values reported show that the effective FoV is enlarged up to a factor $\sim$8 with respect to the parallel mode. The div values chosen allow to span from the maximum value of m$_{ave}$, obtained in parallel mode, to the value chosen set as a threshold (m$_{ave}\sim$ 3).\\
\begin{figure}[h!]
\centering
{\includegraphics[width=5cm]{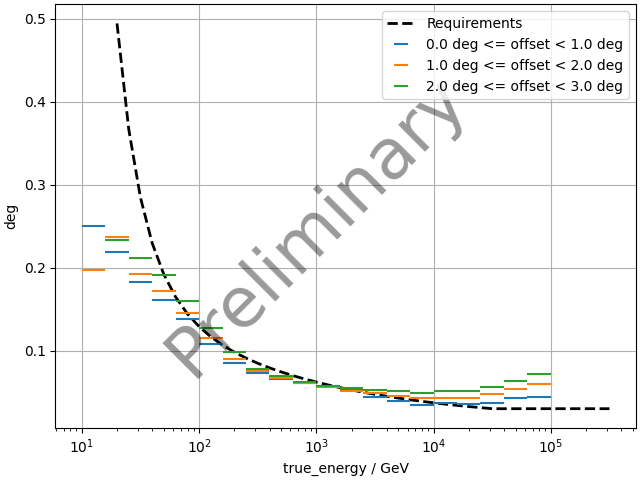}}
{\includegraphics[width=5cm]{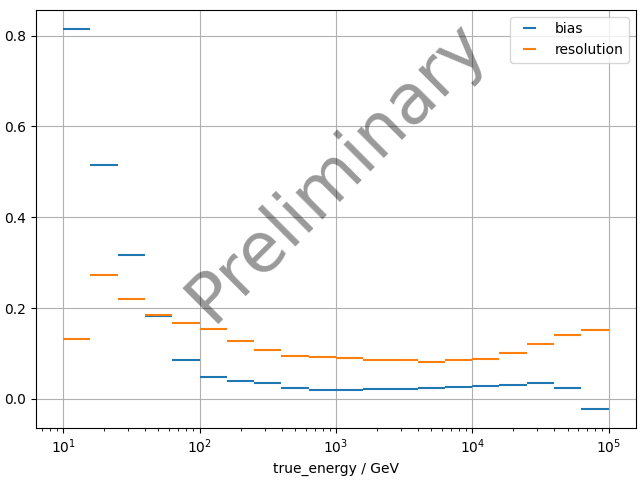}}
\caption{\textbf{Left:} Preliminary angular resolution for cfg2. \textbf{Right:} Preliminary energy bias and resolution for cfg2}
\label{fig:reco}
\end{figure}
    The simulations have been analysed using \texttt{ctapipe}, the python package developed for the processing of CTA low-level data \cite{ref:ctapipe_icrc2021} \cite{ref:ctapipe_zenodo}. The reconstruction method for divergent mode is already available in \texttt{ctapipe} \cite{ref:div_icrc}. From the image recorded by each telescope a plane is defined by the projection of the shower axis on the camera and the telescope position. Those planes, belonging to a 3D reference frame common to all telescopes, are then intersected pair-wise and the angle between them is used as a weight for the computation of the final reconstructed direction. This is computed as a weighed average between all pair-wise directions. The strength of this method is that no correction is needed in the direction reconstruction with respect to the parallel pointing case.
\section{Results and discussion}

\begin{figure}[h!]
    \centering
    \includegraphics[scale=0.35]{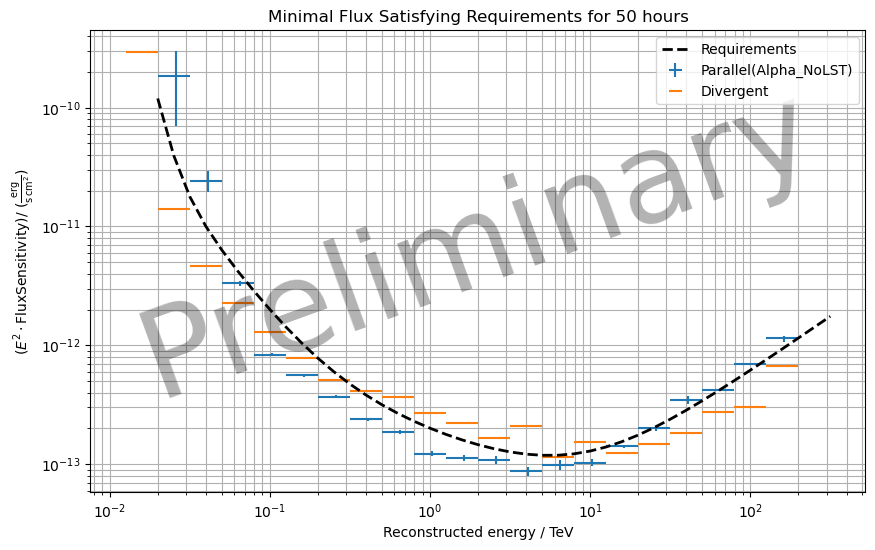}
    \caption{Sensitivity curve for the divergent configuration named cfg2, compared with CTA requirements (black line) and southern array parallel pointing sensitivity (orange). The latter is referred to alpha configuration, where no LSTs are included.}
    \label{fig:sens}
\end{figure}

The goal of this study is to determine some criteria, aside from FoV extension, to constrain the usable divergent configurations. The obvious condition is the array performance: sensitivity, effective area, angular resolution, gamma-hadron separation, point-spread function, ecc. Among possible criteria there are the number of triggered and reconstructed telescopes at various energies and the number of truncated images we obtain for the different configurations. Both studies are at the moment in a preliminary phase. 
The results reported  here refer to a point-like gamma-ray sources. The performance for diffuse sources will also be computed but in this case some of the symmetry assumptions that are valid for parallel pointing might not be satisfied anymore when dealing to divergent pointing. This means that radial symmetry in the FoV must be checked for effective area, energy dispersion and PSF.
The performance presented here is obtained for the configuration named \texttt{cfg2}, see Table \ref{table:divergent_details}. As can be seen from 
\figurename{ \ref{fig:reco}} and \figurename{ \ref{fig:sens}} the performance of this mildly divergent configuration is in line with CTAO requirements. This is a promising result, telling us that, at least up to the divergent value here analyzed, the expected drop in the array performance is not that intense. This condition confirms that divergent pointing is a suitable observational strategy. The expected behaviour for more divergent configurations is a growing worsening in the performance. Since the pipeline for data analysis has been recently changed only one configuration has been analysed at the moment, in order to make sure that everything is working properly code-wise. The analysis and performance plots of the other configurations will be produced soon.  

\section{Conclusions}

Divergent pointing is a promising pointing strategy for the Cherenkov Telescope Array Observatory. Its main objective is to increase the array instantaneous FoV thus reducing the time needed to cover a large area of the sky and increasing the probability to detect transient sources. The drawback of the technique are a reduced angular sensitivity and energy reconstruction capability of the array. The goal is thus to find a set of configurations that allow to exploit the enlarged FoV while maintaining an acceptable performance. With this study the performance of CTAO-South has been analysed for some preliminary configurations. This analysis allowed to test the analysis pipeline, which changed recently and to look into the performance of the array not only at the center of the FoV but also for several offset positions. Only one of the simulated configurations has been analysed at the moment, in the next months the other configurations will be analysed as well.\\ The results obtained so far are promising, since the performance is consistent with the requirements and we don't observe a significant worsening in the array sensitivity. \\ The configurations simulated so far have been selected to better understand the applicability and limits of the method. Next step will be to optimize the configurations to specific science goals.

\section{Acknowledgments}
This work was conducted in the context of the CTA Consortium and CTA Observatory.
We gratefully acknowledge financial support from the agencies and organizations listed here:
https://www.cta-observatory.org/consortium\_acknowledgments/

%
%
%


\begin{thebibliography}{99}
\bibitem{ref:cta}
R. Zanin, \emph{Cherenkov Telescope Array: the World’s largest VHE gamma-ray observatory}, 37th International Cosmic Ray Conference, Berlin, Germany, 2021 Online, published March 18, 2022. \href{https://pos.sissa.it/cgi-bin/reader/conf.cgi?confid=395}{PoS(ICRC2021)005} 

\bibitem{ref:dubus2013}
G. Dubus, \emph{Surveys with the Cherenkov Telescope Array}, Astropart Phys, 43, 317-330, 2013.

\bibitem{ref:Szanecki2015}
Szanecki, M., Sobczy{\'n}ska, D., Nied{\'z}wiecki, A., Sitarek, J., Bednarek, W., \emph{Monte Carlo simulations of alternative sky observation modes with the Cherenkov Telescope Array}, Astroparticle Physics, 67, 33-46, 2015.

\bibitem{ref:PhD_th_Donini}
Donini.A, \emph{Monte Carlo simulation and data analysis of sky observation mode with the Cherenkov Telescope Array}, PhD Thesis, University of Udine, 2020.

\bibitem{ref:prod3}
K. Bernl\"ohr, G. Maier, Gernot and A. Moralejo,
\emph{CTAO Simulation Telescope Models for CORSIKA and sim\_telarray - prod3b}, 2022, doi: 10.5281/zenodo.6219128, \href{https://doi.org/10.5281/zenodo.6219128}{Zenodo}

\bibitem{ref:divtel}
T. Vuillaume, A. Donini and T. Gasparetto, \emph{cta-observatory/divtel:v0.1}, 2022, doi:10.5281/zenodo.6415138,  \href{https://doi.org/10.5281/zenodo.6415138}{Zenodo}

\bibitem{ref:corsika}
D. Heck, J. Knapp, J. N. Capdevielle, G. Schatz and T. Thouw, 
\emph{CORSIKA: A Monte Carlo code to simulate extensive air showers}, FZKA-6019, 1998

\bibitem{ref:prod5}
Cherenkov Telescope Array Observatory and Cherenkov Telescope Array Consortium, \emph{CTAO Instrument Response Functions - prod5 version v0.1}, 2021, doi: 10.5281/zenodo.5499840, \href{https://doi.org/10.5281/zenodo.5499840}{Zenodo}

\bibitem{ref:sim_details}
O. Gueta, \emph{The Cherenkov Telescope Array: layout, design and performance}, Proceedings, 37th International Cosmic Ray Conference, Berlin, Germany, 2021, Online, doi: 10.22323/1.395.0885, \href{https://arxiv.org/pdf/2108.04512}{ arXiv:2108.04512}

\bibitem{ref:ctapipe_icrc2021}
M. Linhofff et al, \emph{ctapipe – Prototype Open Event Reconstruction Pipeline for the Cherenkov Telescope Array}, Proceedings, 38th International Cosmic Ray Conference, Nagoya, Japan

\bibitem{ref:ctapipe_zenodo}
K. Kosack et al, \emph{cta-observatory/ctapipe:v0.19.3}, 2023, doi:10.5281/zenodo.8063139,\\
\href{https://doi.org/10.5281/zenodo.8063139}
{Zenodo}
\bibitem{ref:div_icrc}
A.Donini, T.Gasparetto, J.Bregeon, F.Di Pierro,  F.Longo, G.Maier, A.Moralejo, T.Vuillaume, \emph{The Cherenkov Telescope Array Performance in Divergent Mode}, Proceedings, 36th International Cosmic Ray Conference, doi: 10.22323/1.358.0664, \href{https://arxiv.org/abs/1907.07978}{arXiv:1907.07978}




\end{thebibliography}
\end{document}